\documentclass[10pt,journal,compsoc]{IEEEtran}

\usepackage{amsmath}
\usepackage{booktabs}
\usepackage{graphicx}
\usepackage{xcolor}
\usepackage{hyperref}
\usepackage{multirow}

\begin{document}

\title{Trillion-Parameter MoE in a Box: Decoupling Memory Provisioning with High-Bandwidth Flash}

\author{
Pengfei~Xia, Tuo~Hao, Shengwei~Li, Jinjing~Chen, Shiru~Wei, Wenjun~Zou, Rui~Zhang, and~Hui~Zang%
\thanks{The authors are with Huawei Technologies Co., Ltd.}%
}

\IEEEtitleabstractindextext{
\begin{abstract}
An MoE appliance for trillion-parameter models at low concurrency must host terabytes of weights on one node and serve prefill and decode with fixed resources. Combining operator analysis of two trillion-parameter MoE models, a measured expert routing trace, and agentic serving traces over multiple turns, we explore a design space spanning High-Bandwidth Flash (HBF) and DRAM configurations, bandwidth exposure, and near-data compute. We find that state bandwidth and HBF transport form two largely orthogonal knees and address two provisioning questions. Q1: Once weights move to HBF, what bandwidth-to-capacity ratio does DRAM require? With a 256-GB floor for the state tier, both models meet a $1.10\times$ completion time target at ratios of only $1.4$--$4.0~\mathrm{s}^{-1}$, roughly an order of magnitude below HBM3e's $33.3~\mathrm{s}^{-1}$. Q2: As HBF internal bandwidth scales with capacity, must bandwidth to the host scale proportionally? Six HBF packages expose 384~GB/s per package to the host, 2.30~TB/s aggregate and 62.5\% below the 6.14~TB/s full exposure reference, while more packages reduce required bandwidth per package at the same target.
\end{abstract}

}

\maketitle
\IEEEdisplaynontitleabstractindextext
\IEEEpeerreviewmaketitle

\section{Introduction}
\IEEEPARstart{D}{atacenter} LLM serving commonly relies on continuous batching and Prefill--Decode disaggregation so that large request pools can amortize model residency and independently provision the two phases \cite{zhong2024distserve}. We study a different deployment regime: \emph{a compact, low-concurrency trillion-parameter MoE appliance}. The entire model must remain resident in one node, concurrency is typical $1$--$8$, and cold prefill, cache-hit incremental prefill, speculative verification, and decode must share one fixed resource set.

In this regime, the first-order challenge is model residency. Quantized trillion-parameter MoE weights occupy TB scale \cite{xu2026deepseek,team2026kimi}, making an HBM-only appliance difficult to scale within a compact system. High-Bandwidth Flash (HBF) provides a high density tier well matched to the read-dominant access pattern of model weights, offering hundreds of GB per package and TB/s-class read bandwidth \cite{ocp2026hbf,ha2026h,wang2026flashaccel}. HBF makes trillion-parameter residency practical in a compact node; once weights are separated from runtime state, however, DRAM provisioning and HBF bandwidth delivery must both be reconsidered.

An HBM-centric organization couples resources driven by different workload behaviors: weight capacity by model size, weight delivery by expert activation and reuse, and runtime-state demand by context length and attention behavior. Once weights move to HBF, DRAM no longer scales with model residency. And as model scale grows, more HBF packages host weights and their aggregate internal bandwidth grows, so proportional exposure would make host I/O scale with model size as well.

This separation leads to two provisioning questions. \textbf{Q1: Once weights move to HBF, what bandwidth-to-capacity ratio does the state-storage DRAM require?} \textbf{Q2: As HBF internal bandwidth scales with capacity, must host-facing bandwidth scale proportionally?} Prior HBF studies cover device characterization, HBM--HBF placement, system integration, and direct HBF data paths \cite{son2026exploring,ha2026h,wang2026flashaccel,kim2026dash}. This work quantifies the state-tier and host-transport provisioning knees that emerge after model weights reside in HBF.

We explore a design space covering HBF and DRAM configurations, HBF bandwidth exposure, and near-data compute. With model weights in HBF, both models meet a $1.10\times$ completion time target at DRAM bandwidth-to-capacity ratios of $4.0$ and $1.4~\mathrm{s}^{-1}$ for DSV4-Pro and Kimi-K3, respectively, well below HBM3e's $33.3~\mathrm{s}^{-1}$. An LPCAMM2$\times8$ tier reaches the required bandwidth at $1.09~\mathrm{TB/s}$, whereas similar $256$-GB configurations fall short for DSV4-Pro. For HBF, six packages remain within $10\%$ of the high bandwidth reference with $384~\mathrm{GB/s}$ exposed per package, or $2.30~\mathrm{TB/s}$ in aggregate, $62.5\%$ below the $6.14~\mathrm{TB/s}$ reference. More packages increase aggregate internal bandwidth and reduce required host bandwidth per package at the same target.

\section{Workload and Design Space}
We use DSV4-Pro \cite{xu2026deepseek} and Kimi-K3 \cite{team2026kimi} as representative trillion-parameter MoE workloads. As summarized in Table~\ref{tab:dsv4_kimik3}, both models combine TB-scale weights with sparse expert activation and long-context attention, but differ substantially in MoE and attention organization.

\begin{table}[t]
\centering
\caption{Model architecture and memory characteristics.}
\label{tab:dsv4_kimik3}
\resizebox{1.0\columnwidth}{!}{
\begin{tabular}{c|c|c}
\toprule
\textbf{Metric} & \textbf{DSV4-Pro} & \textbf{Kimi-K3} \\ \midrule
Total / activated parameters & $1.6$T / $49$B & $2.8$T / $104$B \\
Layers & $61$ & $93$ \\
Routed experts & $384$, Top-$6$ & $896$, Top-$16$ \\
Expert core dims & $7168 \rightarrow 3072 \rightarrow 7168$ & $3584 \rightarrow 3072 \rightarrow 3584$ \\
Attention & $30$ CSA + $31$ HCA & $69$ KDA + $24$ Gated MLA \\
Quantized weights & $865.0~\mathrm{GB}$ & $1560.0~\mathrm{GB}$ \\
State, $1$M tokens, $B{=}1$ & $4.66~\mathrm{GB}$ & $28.99~\mathrm{GB}$ \\
\bottomrule
\end{tabular}}
\end{table}

\noindent \emph{Weight and runtime state have different scaling behavior.} At a $1$M-token context, persistent state occupies only $4.66~\mathrm{GB}$ for DSV4-Pro and $28.99~\mathrm{GB}$ for Kimi-K3. At concurrency eight, weights account for $95.9\%$ and $87.1\%$ of total residency. The TB-scale capacity requirement of an appliance is thus imposed primarily by model weights; the DRAM left after weight migration serves a smaller dynamic-state working set. This separation motivates Q1: the bandwidth-to-capacity ratio appropriate for state storage need not inherit that of the memory selected to host the model.

\begin{figure}[t]
\centering
\includegraphics[width=1.0\linewidth]{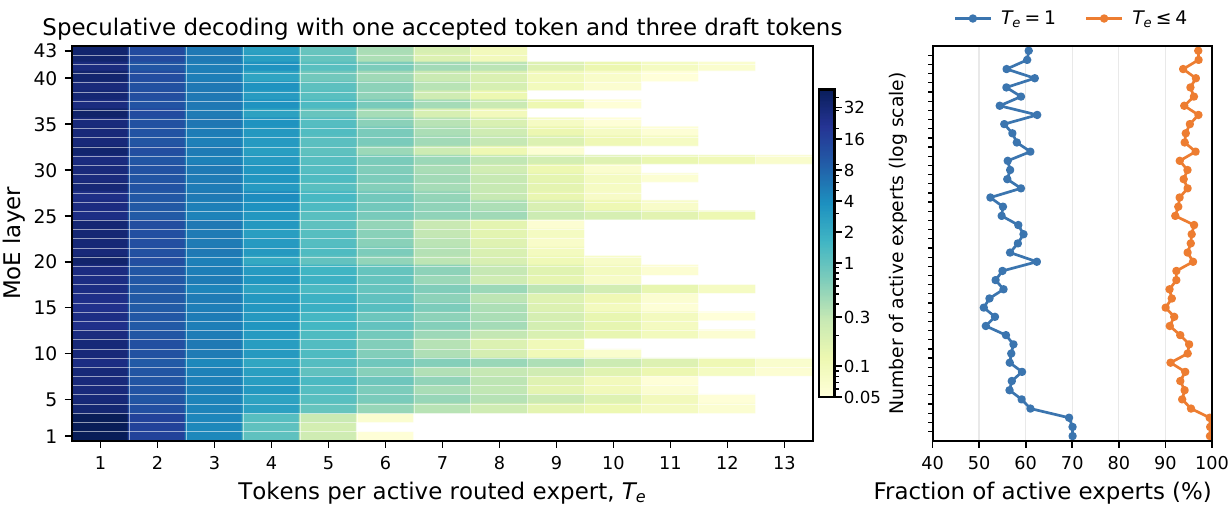}
\caption{Per-layer routed-expert occupancy of DSV4-Flash \cite{xu2026deepseek}.}
\label{fig:expert_occupancy}
\end{figure}

\noindent \emph{Sparse routing makes weight traffic difficult to reuse.} A router trace from a SPEED-Bench code \cite{abramovich2026speed} session shows that $16$ effective tokens activate $51.6$ experts on average, with $57.8\%$ of activated experts receiving one token (Figure~\ref{fig:expert_occupancy}). Routed-expert execution often operates with a small effective $M$ dimension, limiting weight reuse and making decode sensitive to weight delivery, the behavior behind Q2.

\noindent \emph{Dynamic-state traffic is less bandwidth intensive.} Compression and sparsity reduce attention state traffic, while speculative verification increases computation per state read. For example, Kimi-K3's absorbed MLA core reads about $9.7~\mathrm{GB}$ of latent KV per layer at batch size $8$ and a $1$M-token sequence length; increasing effective tokens per verification step from $1$ to $6$ changes state traffic by less than $0.1\%$, while arithmetic intensity rises from $182.3$ to $1092.9~\mathrm{FLOP/B}$.

\noindent \textbf{System model.} We model the appliance as a central compute die connected to a DRAM state tier and multiple HBF packages hosting model weights. The compute die is held fixed at $450~\mathrm{TFLOP/s}$ FP8 throughput with an $M$-dimension partition of $16$ as modeling constants, isolating memory-system effects. The state tier sweeps three representative module points (HBM3e: $36~\mathrm{GB}$, $1.2~\mathrm{TB/s}$ per stack; SOCAMM2: $128~\mathrm{GB}$, $153.6~\mathrm{GB/s}$ per module; LPCAMM2: $64~\mathrm{GB}$, $136.5~\mathrm{GB/s}$ per module), whose bandwidth-to-capacity ratios are $33.3$, $1.20$, and $2.13~\mathrm{s}^{-1}$. Each modeled HBF provides $512~\mathrm{GB}$ and $B_{\mathrm{int}}=1.6~\mathrm{TB/s}$ read bandwidth, with $16$ NAND dies, $96$ planes per die, $4~\mathrm{KB}$ pages, and $t_R=4~\mu\mathrm{s}$ \cite{wang2026flashaccel,ocp2026hbf}, modeling page granularity, die--plane parallelism, and read amplification. Part of $B_{\mathrm{int}}$ is exposed as $B_{\mathrm{ext}}$, defining the exposure ratio $\eta_{\mathrm{exp}}=B_{\mathrm{ext}}/B_{\mathrm{int}}$. We sweep $B_{\mathrm{ext}}$ in \emph{host-link quanta} of $96~\mathrm{GB/s}$ ($1$--$16$ per HBF); at $16~\mathrm{GT/s}$ one quantum equals one x64-equivalent payload group at $75\%$ efficiency. Optionally, each HBF may carry an FP8 engine ($8\times1024~\mathrm{MAC/cycle}$ at $1~\mathrm{GHz}$, $16.8~\mathrm{MB}$ LHB \cite{ha2026h}) for low-use routered experts.

\noindent \textbf{Methodology.} We enumerate DRAM technology and count, HBF count ($\{1,2,4,6,8,10,12\}$), host-link quanta per HBF, and near-HBF compute ($\{\mathrm{Off},\mathrm{On}\}$), removing designs that violate weight capacity, runtime-state capacity, or execution constraints. Completion time denotes the end-to-end completion time of a complete multi-turn conversation. Evaluation uses three multi-turn agentic traces with distinct execution compositions: Trace A is more mixed and prefill-heavy ($32$ turns, $68$K--$261$K-token contexts, $94.1\%$ cache-hit rate), Trace B combines cached contexts with a larger decode share ($29$ turns, $18$K--$106$K-token inputs, $94.6\%$ hit rate), and Trace C has the largest pure-decode fraction ($52$ turns, $18$K--$141$K-token inputs, $96.6\%$ hit rate). Each trace is replicated into $8$ sessions with mean $100$ ms exponential inter-arrival gaps. Serving is modeled with LLMServingSim 2.0 \cite{cho2026llmservingsim}, using continuous batching, chunked prefill beyond $8{,}192$ tokens, and speculative verification with $6$ effective tokens per step and acceptance length $4.27$ \cite{cheng2026dspark}. Routed-expert occupancy is modeled by a Pólya-urn calibrated on the SPEED-Bench traces and instantiated with each model's expert count and Top-$K$.

\begin{figure*}[t]
\centering
\includegraphics[width=1.0\textwidth]{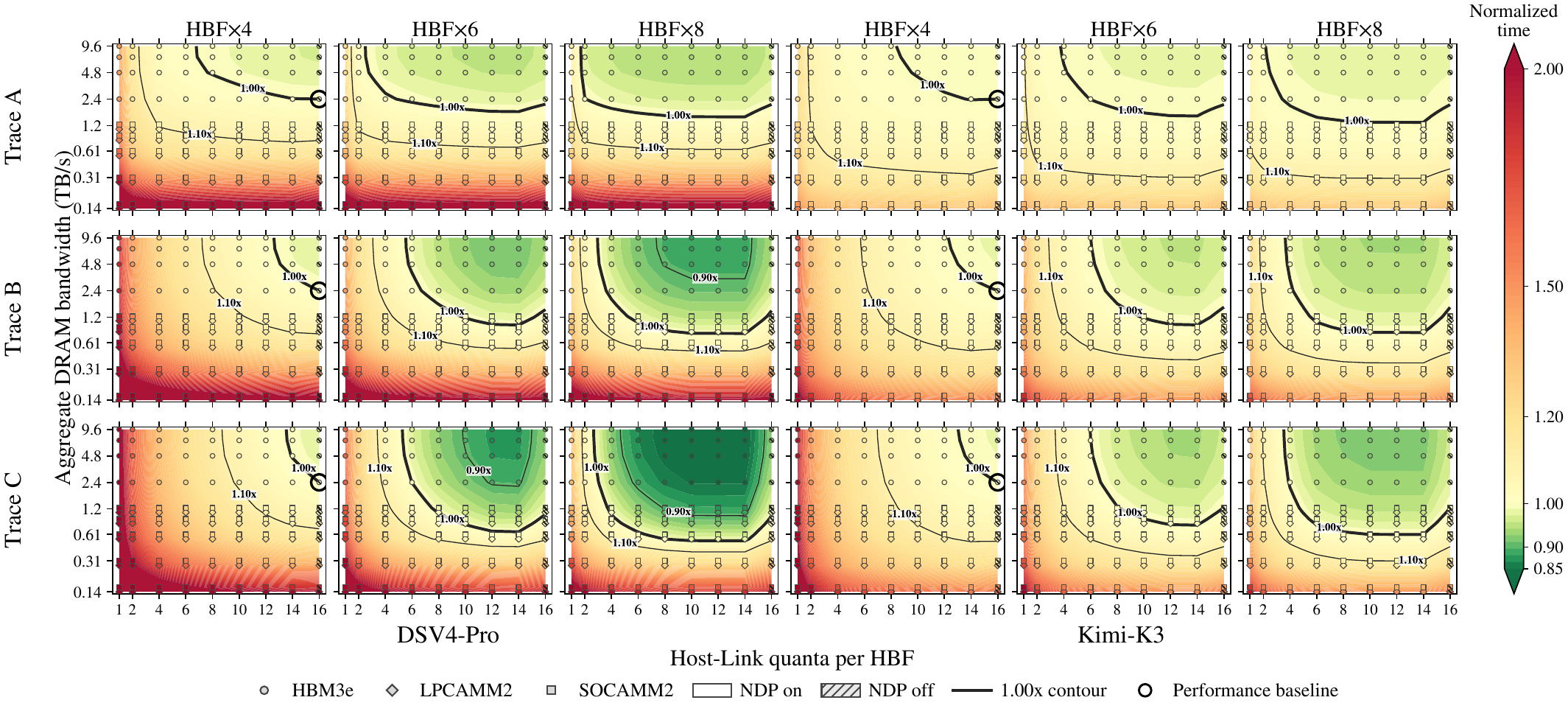}
\caption{Completion-time landscape over state-tier bandwidth and per-HBF host-link bandwidth. Contours are normalized to the circled DRAM-bandwidth-saturated reference.}
\label{fig:dse_results}
\end{figure*}


\section{Evaluation}
All times are normalized to a DRAM-bandwidth-saturated reference, circled in Figure~\ref{fig:dse_results}: HBM3e$\times2$ with HBF$\times4$ at full host exposure and near-HBF execution off. Its $2.4~\mathrm{TB/s}$ aggregate state bandwidth lies in the saturated region, where additional DRAM bandwidth yields little improvement; we use this point to define $1.00\times$ and the $1.10\times$ provisioning contour, a target allowing up to $10\%$ completion-time degradation. The figure spans aggregate DRAM bandwidth (vertical) and per-HBF host-link quanta (horizontal) over both models, three HBF counts, and three traces; markers show realizable HBM3e/LPCAMM2/SOCAMM2 points, with hatching denoting near-HBF execution off. Vertical sweeps expose the state-bandwidth knee at sufficient host exposure; horizontal sweeps expose the HBF host-link knee at sufficient state bandwidth.

The dominant structure in every panel is an \emph{L-shaped} iso-performance landscape. Completion time degrades sharply when \emph{either} resource falls into its severely starved region, roughly one to two host-link quanta per HBF or $0.14$--$0.31~\mathrm{TB/s}$ of DRAM, and the two legs of the L barely interact: raising DRAM bandwidth to $9.6~\mathrm{TB/s}$ cannot rescue a one-quantum design, and opening all sixteen quanta cannot rescue a $0.14~\mathrm{TB/s}$ state tier. We describe the two knees as \emph{largely orthogonal}: after one resource leaves its severely starved region, additional bandwidth in that dimension has little effect on the knee of the other. The goal of provisioning is therefore the lower-left knee of the $1.10\times$ contour, the lowest-resource operating point at which neither bandwidth dimension is substantially over-provisioned.

\subsection{The State-Tier Knee Sits Far Below HBM}
Reading the vertical direction at adequate exposure ($\geq\!8$ quanta), the landscape is steep below $0.61~\mathrm{TB/s}$, moderate up to $1.2~\mathrm{TB/s}$, and nearly flat beyond: the worst-case state-bandwidth threshold for the $1.10\times$ target sits around $1~\mathrm{TB/s}$, while all HBM3e rows ($2.4$--$9.6~\mathrm{TB/s}$) lie deep in the saturated region. An HBM-inherited state tier is thus massively over-provisioned once weights move to HBF, and the LPDDR rows already straddle the knee.

\begin{figure}[t]
\centering
\includegraphics[width=1.0\linewidth]{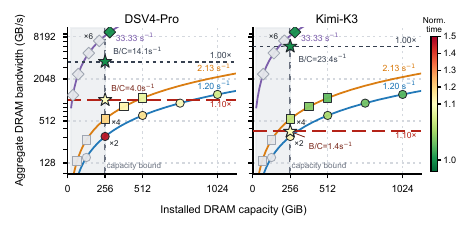}
\caption{State-tier bandwidth--capacity provisioning for Trace A at HBF$\times4$.}
\label{fig:dram_provisioning}
\end{figure}

Figure~\ref{fig:dram_provisioning} quantifies this at HBF$\times4$ on Trace A by placing every feasible module count on its technology's bandwidth--capacity curve. Two constraints structure the space. First, \emph{capacity}: the first feasible module-count point lies in the $256$-GB class (shaded bound), which excludes HBM3e$\times6$ ($216~\mathrm{GB}$) despite its $7.2~\mathrm{TB/s}$. It is set by the stricter Kimi-K3 case, where a $1$M-token sequence occupies about $29~\mathrm{GB}$ of persistent state and eight concurrent sequences require about $232~\mathrm{GB}$ before serving overhead. Second, \emph{performance}: at the $256$-GB provisioning floor, the $1.10\times$ performance boundary corresponds to bandwidth-to-capacity ratios of $4.0~\mathrm{s}^{-1}$ for DSV4-Pro and $1.4~\mathrm{s}^{-1}$ for Kimi-K3 (star markers), versus $33.3~\mathrm{s}^{-1}$ for HBM3e, an $8\times$--$24\times$ relaxation.

Because each device has a fixed bandwidth-to-capacity ratio $r_{\mathrm{dev}}$, bandwidth demand translates into excess capacity: $C_{\mathrm{install}}\geq\max(C_{\min},B_{1.10}/r_{\mathrm{dev}})$. LPCAMM2$\times4$ and SOCAMM2$\times2$ meet the capacity floor, yet for DSV4-Pro they remain outside the $1.10\times$ line (SOCAMM2$\times2$ at $\approx\!1.5\times$): their ratios of $2.13$ and $1.20~\mathrm{s}^{-1}$ undershoot the required $4.0~\mathrm{s}^{-1}$ at that capacity, so LPCAMM2 reaches the boundary only at $\times8$ ($512~\mathrm{GB}$, $1.09~\mathrm{TB/s}$). Kimi-K3 requires less bandwidth: LPCAMM2$\times4$ already sits at the boundary, and both $\times8$ LPDDR configurations lie within it. The state tier is sized by the knee, capacity first and then aggregate bandwidth; the ratio of the weight-hosting memory no longer enters the decision.

\subsection{Host Bandwidth Need Not Track Internal Bandwidth}
Reading the horizontal direction, time improves steeply from one to two to four quanta per HBF, the knee sits at four to eight quanta depending on model and trace, and from twelve to sixteen quanta the contours are flat: exposing the full $1.6~\mathrm{TB/s}$ internal bandwidth is unnecessary. Concretely, at HBF$\times6$ with four quanta per HBF each package exposes $4\times96=384~\mathrm{GB/s}$, $\eta_{\mathrm{exp}}=0.24$, or $2.30~\mathrm{TB/s}$ in aggregate, $62.5\%$ below the $6.14~\mathrm{TB/s}$ full-exposure reference ($4$ HBF $\times$ $16$ quanta per HBF $\times$ $96~\mathrm{GB/s}$ per quantum); yet with an LPDDR-class state tier both models remain at or inside the $1.10\times$ boundary across traces.

\emph{HBF count is itself a performance dimension.} From HBF$\times4$ to $\times6$ to $\times8$, the green region expands toward the lower-left: completion time is reached with fewer quanta per package. HBF$\times4$ at sixteen quanta and HBF$\times8$ at eight quanta deliver the same $6.14~\mathrm{TB/s}$ host-facing aggregate, yet on Trace C for DSV4-Pro the former remains at $\approx\!1.05$--$1.10\times$ while the latter enters the $0.90\times$ region; the packages contribute internal read bandwidth, die--plane parallelism, and near-HBF compute that no per-link widening of four packages can replicate. Capacity alone requires HBF$\times2$ for DSV4-Pro ($865~\mathrm{GB}$) and HBF$\times4$ for Kimi-K3 ($1560~\mathrm{GB}$, beyond three packages' $1536~\mathrm{GB}$), yet capacity-minimal provisioning, $N_{\mathrm{HBF}}=\lceil \text{model size}/\text{package capacity}\rceil$, leaves performance on the table.

\subsection{The Knees Are Model- and Trace-Dependent}
The two models exhibit different state-bandwidth landscapes because their attention organization, MoE structure, and runtime operator composition differ. On Trace A, the $1.10\times$ contour lies at approximately $0.6$--$1.2~\mathrm{TB/s}$ for DSV4-Pro and $0.4$--$0.6~\mathrm{TB/s}$ for Kimi-K3, corresponding to $4.0$ and $1.4~\mathrm{s}^{-1}$ at the common $256$-GB state-capacity floor. Prefill is Cube-dominated ($63$--$74\%$ of operator time) for both workloads, whereas decode contains different fractions of fused attention and Cube kernels: Kimi-K3 shifts toward long fused attention ($\approx\!48\%$ Fused versus $\approx\!50\%$ Cube) while DSV4-Pro remains $\approx\!86\%$ Cube. Speculative verification further amortizes attention-state reads across multiple effective tokens. A smaller fraction of Kimi-K3's makespan is therefore sensitive to central DRAM bandwidth, matching its flatter contours.

\begin{figure}[htbp]
\centering
\includegraphics[width=1.0\linewidth]{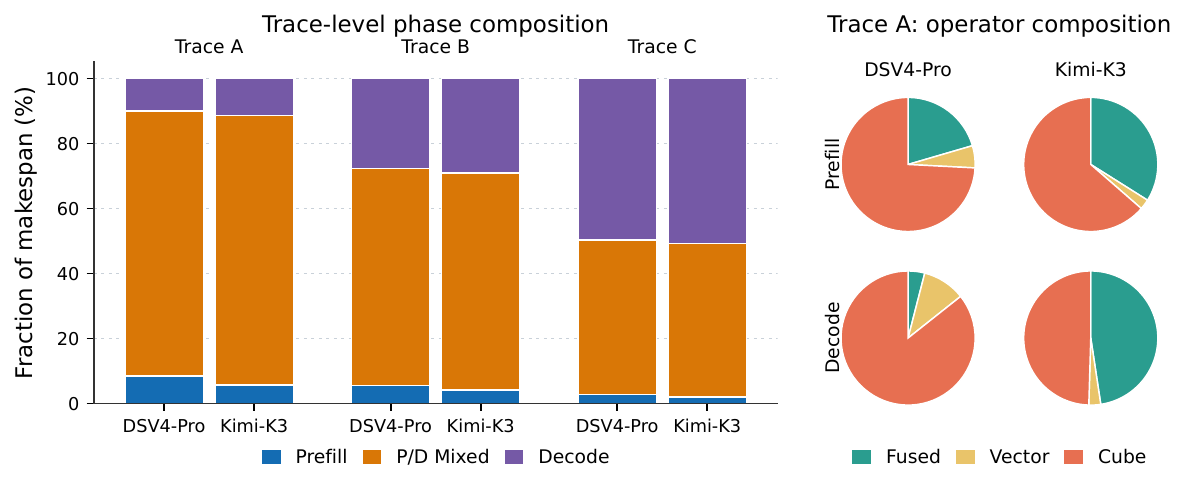}
\caption{Workload composition across traces and operators.}
\label{fig:workload_composition}
\end{figure}

The landscape is equally trace-dependent. The pure-decode share of makespan grows from $\approx\!11\%$ (Trace A) to $\approx\!28\%$ (Trace B) and $\approx\!50\%$ (Trace C), and because mixed phases batch many tokens per activated expert while pure decode strains weight delivery, the link knee shifts rightward and steepens from Trace A to Trace C: configurations that sit comfortably inside $1.10\times$ on Trace A fall outside it on Trace C, while HBF-rich designs on Traces B and C open sub-$1.00\times$ regions ($0.90\times$) against the capacity-stressed reference. The traces differ simultaneously in context length, cache-hit rate, and phase composition, and context length alone does not explain these shifts. The provisioning knee thus depends on end-to-end phase composition in addition to context length and cache-hit behavior.

\subsection{Packaging Implications}
\label{sec:packaging}
The two bandwidth knees reduce pressure on package I/O. The state tier can be provided by board-level LPCAMM2 or SOCAMM2 modules; an LPCAMM2$\times8$ tier supplies the required $512~\mathrm{GB}$ and $1.09~\mathrm{TB/s}$ (SOCAMM2$\times8$: $1{,}024~\mathrm{GB}$, $1.23~\mathrm{TB/s}$), removing state capacity and bandwidth from compute-die shoreline competition. On the HBF side, the $1.10\times$ operating point requires only partial exposure of each package's internal bandwidth, and the per-package link width falls further as HBF capacity scales.

UCIe-S provides a standard-package integration path on organic substrates, while UCIe-A provides higher link density for advanced packaging \cite{ucie2025spec}. The reduced host-exposure requirement moves the HBF appliance toward a regime that can be served by lower-density package links. Final package feasibility still depends on die placement, routing, reach, signal integrity, and power delivery. The design-space exploration therefore shows that host-I/O bandwidth density need not scale proportionally with HBF capacity, relaxing one of the key pressures that motivates advanced packaging.

\section{Conclusion}
HBF changes more than model capacity. Once weights and runtime state are separated, three resources can follow three workload drivers: HBF capacity follows resident weights, state bandwidth follows runtime state, and host transport follows activated-expert traffic. State bandwidth and HBF host transport exhibit largely separable performance knees, so the two resources can be provisioned independently once severe starvation is avoided. Across two trillion-parameter models, a common $256$-GB state tier needs only $1.4$--$4.0~\mathrm{s}^{-1}$, and six HBF packages expose only $384~\mathrm{GB/s}$ each at a $1.10\times$ target; added packages further reduce per-package exposure. Partial exposure in turn relaxes package-I/O pressure, making standard organic-substrate packaging a viable path for a trillion-parameter appliance.

{\footnotesize
\bibliographystyle{IEEEtran}
\bibliography{refs}
}

\end{document}